\documentclass[twocolumn,fleqn]{article}
\usepackage{amsmath,amssymb,amsfonts}
\usepackage{graphicx}						
\usepackage{url}
\title{A nucleation and growth model for COVID-19 epidemic in Japan}
\author{Yoshihiko Takase\thanks{Chiba JICA Senior Volunteers Association}}

\begin{document}
\maketitle

\abstract{}
COVID-19 epidemics in Japan and Tokyo were analyzed by a fundamental equation of the dynamic phase transition. As a result, the epidemic was found to be in good agreement with the random nucleation and linear growth model suggesting that the epidemic between March 13, 2020 and May 22, 2020 was simply rate-limited by the three constant-parameters: the initial susceptible, domain growth rate, and nucleation decay constant. This model provides a good predictor of the epidemic because it consists of one equation and the initial specific plot is linear.

\section{Introduction}
In Japan, the first case of COVID-19 was reported on January 16, 2020. The Japanese government admitted the cruise ship "Diamond Princess" with the patients to the port on February 3, and began quarantine\cite{diamond_princess}. This news was widely reported and provided an opportunity for most Japanese to know the danger of COVID-19. On February 24, 2020, the National Expert Meeting announced its views\cite{expert_meeting} and on February 27, Prime Minister Abe requested schools all over the country to close\cite{school_close}. The government announced a state of emergency to seven prefectures on April 7, 2020, and expanded it nationwide on April 16\cite{emergency}. The daily number of new infections in Japan reached a peak around April 12th, and then began to decline, reaching around 30 in late May.

We previously analyzed the ferroelectric polarization reversal phenomenon of polymers\cite{ferro_pol_rev} by a nucleation and domain growth model of the dynamic phase transition, one of the general theory of physics. This time, we have noted that the polymers, which are complex systems of crystalline and amorphous phases, resemble human society and that the time dependence of COVID-19 new infections resembles the polarization reversal characteristics. The purpose of this study is to see if the fundamental equation of the model is directly applicable to the spread of COVID-19 infection, although the epidemic is commonly analyzed by the SIR and its expanded models\cite{covid19_sir_model,covid19_stochastic_sir_model,covid19_shir_model}.

\section{Theoretical basis}

We first consider the fraction of domain $X'$ that has been transformed by formation and growth of fictitious nuclei without mutual impingement. This fraction $X'$ at time $t$ is expressed in the following form\cite{leo_mandelkern},
\begin{equation} \label{eq:massfr_x_t_avrami}
	X'  = \int_{0}^{t} v(t, \tau) \dot{N}(\tau) \mathrm{d}\tau,
\end{equation}
where $v(t, \tau)$ is the volume of a nucleus born at time $\tau$ and grown without any restriction until time $t$ $(\tau \leq t)$ and $\dot{N}(\tau) $ is the nucleation probability per unit volume of untransformed region. The term in the integrand involves the usual random nucleation at a rate of $\dot{N}(\tau) = N_0 \nu \mathrm{e}^{-\nu \tau}$
 followed by a steady-state domain growth, where $N_0$ is the number of active points for nucleus and $\nu$ the decay constant. 

The actual volume fraction $X$ that has undergone transformation at time $t$ is related to $X'$ by\cite{avrami}
\begin{equation} \label{eq:avrami_eq}
	\frac{\mathrm{d}X}{\mathrm{d}X'} = 1 - X.
\end{equation}

Integrating Eq.(\ref{eq:avrami_eq}), we obtain
\begin{equation} \label{eq:int_avrami_eq}
	X = 1 - \exp(-X').
\end{equation}

Substituting $X'$ in Eq.(\ref{eq:massfr_x_t_avrami}) into Eq.(\ref{eq:int_avrami_eq}), we obtain the following fundamental equation of the dynamic phase transition.
\begin{equation} \label{eq:avrami_sol_x}
	X = 1 - \exp \left[- \int_{0}^{t} v(t, \tau) \dot{N}(\tau) \mathrm{d}\tau \right].
\end{equation}

If we assume the total number of daily new infections $D(t)$ to be proportional to the transformed volume fraction $X$, then
\begin{equation} \label{eq:avrami_sol_d}
	D(t) = D_s\left[1 - \exp \left(- \int_{0}^{t} v(t, \tau) \dot{N}(\tau) \mathrm{d}\tau \right)\right],
\end{equation}
where $D_s$ is the initial susceptible $D(\infty)$.

\subsection{Random nucleation and one-dimensional linear growth}

The volume that is born at time $\tau$ and grows one-dimensionally until time $t$ $(\tau \leq t)$ is
\begin{equation} \label{eq:volume1dim}
	v(t, \tau) = S_c G (t-\tau),
\end{equation}
where $G$ is the growth speed and $S_c$ is the growth cross section.

Integrating Eq.(\ref{eq:avrami_sol_x}) after substitution of Eq.(\ref{eq:volume1dim}),
\begin{equation} \label{ln1div1_x_sol_1dim}
	\ln\frac{1}{1-X} = -\frac{S_c G N_0}{\nu} f_1,
\end{equation}
where $f_1 = 1 - \mathrm{e}^{-\nu t} - \nu t$.

When $\nu t$ is small enough, Eq.(\ref{ln1div1_x_sol_1dim}) becomes 
\begin{equation} \label{ln1div1_x_sol_s_nt_1dim}
	\ln \frac{1}{1-X} \simeq \frac{S_c G N_0 \nu}{2} t^{2}.
\end{equation}

The total number of daily new infections is
\begin{equation} \label{eq:avrami_sol_d_1dim}
	D(t) = D_s \left[ 1 - \exp\left( \frac{S_c G N_0}{\nu} f_1 \right) \right].
\end{equation}

The number of daily new infections is driven by differentiation of Eq.(\ref{eq:avrami_sol_d_1dim}) as
\begin{equation} \label{eq:avrami_sol_j_1dim}
	J(t) = D_s S_c G N_0(1 - \mathrm{e}^{-\nu t}) \exp\left(\frac{S_c G N_0}{\nu} f_1 \right).
\end{equation}

\subsection{Random nucleation and two-dimensional linear growth}
The volume for the two-dimensional growth is
\begin{equation} \label{eq:volume2dim}
	v(t, \tau) = \pi G^{2} (t-\tau)^{2} l_c,
\end{equation}
where $l_c$ is the domain thickness.

Integrating Eq.(\ref{eq:avrami_sol_x}) after substitution of Eq.(\ref{eq:volume2dim}),
\begin{eqnarray} \label{ln1div1_x_sol_2dim}
	\ln\frac{1}{1-X} = \frac{2\pi G^{2}l_c N_0}{\nu^{2}} f_2,
\end{eqnarray}
where $f_2 = 1 - \mathrm{e}^{-\nu t} - \nu t + (\nu t)^2 / 2$.

When $\nu t$ is small enough, Eq.(\ref{ln1div1_x_sol_2dim}) becomes 
\begin{equation} \label{ln1div1_x_sol_s_nt_2dim}
	\ln\frac{1}{1-X} \simeq \frac{\pi G^{2}l_c N_0 \nu}{3} t^3.
\end{equation}

The total number of daily new infections is
\begin{equation} \label{eq:avrami_sol_d_2dim}
	D(t) = D_s \left [ 1 - \exp \left ( -\frac{2\pi G^2 l_c N_0}{\nu^2} f_2 \right ) \right ].
\end{equation}

The number of daily new infections is
\begin{equation} \label{eq:avrami_sol_j_2dim}
	J(t) = D_s \left( -\frac{2\pi G^2 l_c N_0}{\nu} f_1 \right) \exp\left(-\frac{2\pi G^2 l_c N_0}{\nu^2} f_2 \right).
\end{equation}

\section{Data analysis}
\subsection{COVID-19 data of Japan}

\begin{figure} [p]
 \centering
  \includegraphics[width=8.0cm]{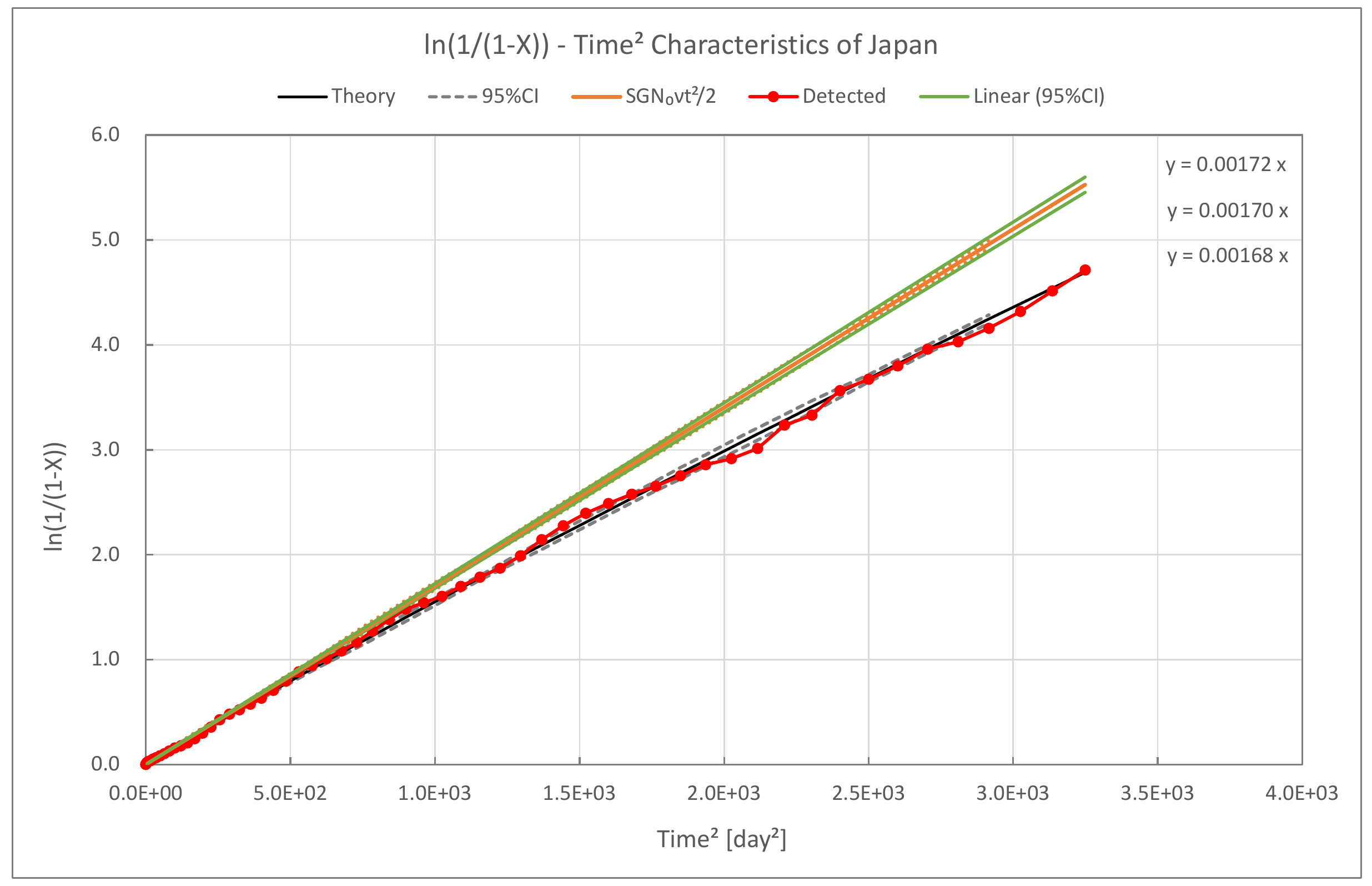}
  \caption{$\mathrm{ln}(1/(1-X)) – Time^2$ characteristics of Japan. The red marker represents the detected infections, the black line the theoretical curve (1-dim.), the gray broken lines the 95\%CI, the orange line the theoretical line (1-dim.) for the small $\nu t$ region, and the green lines linear approximations of the 95\%CI curves.}
  \label{fig:ln1div1_x_t2_japan}
\end{figure}

The COVID-19 data of Japan is provided in some web sites. Using the data\cite{infections_japan}, and assuming the one-dimensional growth, Fig. \ref{fig:ln1div1_x_t2_japan} shows the $\mathrm{ln}(1/(1-X)) – Time^2$ characteristics of Japan. The red marker is the detected infections, the black line is the theoretical curve of Eq.(\ref{ln1div1_x_sol_1dim}), and the orange line is Eq.(\ref{ln1div1_x_sol_s_nt_1dim}). 

The 95\% confidence interval (95\%CI) was obtained by the moving standard deviation calculated over a sliding window of 7 days across neighboring days. The gray broken lines neighboring the theoretical line represent the 95\%CI. Appling the same 95\%CI data to Eq.(\ref{ln1div1_x_sol_s_nt_1dim}), the slope of the orange line was estimated to be $0.00170 \pm 0.00002$ as shown in the figure, which gives the value of $S_c G N_0 \nu / 2$ in Eq.(\ref{ln1div1_x_sol_s_nt_1dim}).

\begin{figure} [p]
  \centering
  \includegraphics[width=8.0cm]{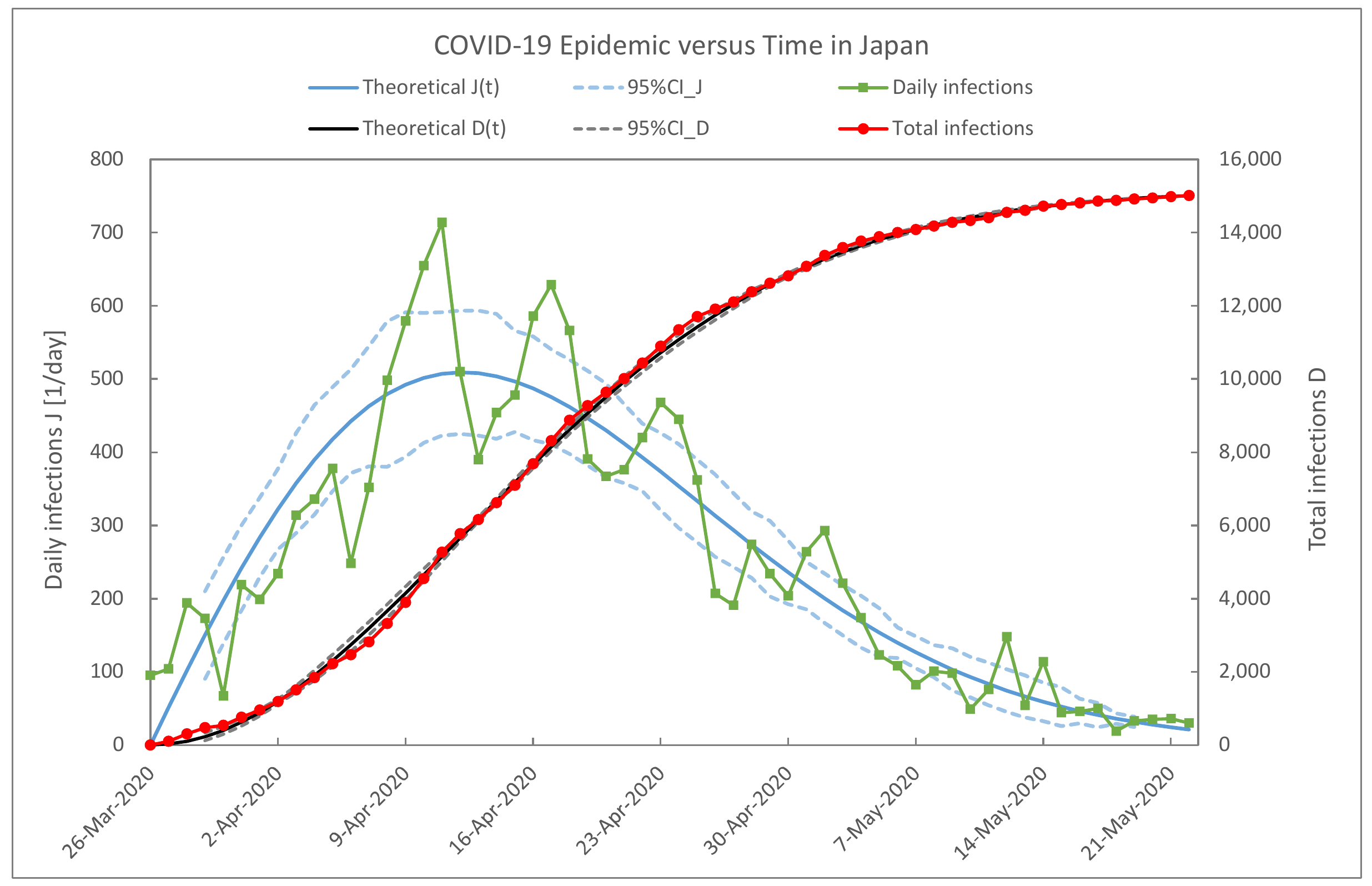}
  \caption{COVID-19 epidemic versus time in Japan. The red marker represents the detected total infections, the black line the theoretical curve (1-dim.) of $D(t)$, the gray broken lines the 95\%CI, the green marker the detected daily infections, the blue curve the theoretical curve (1-dim.) of $J(t)$, and the blue broken lines the 95\%CI curves.}
  \label{fig:infections_japan_1dim}
\end{figure}

Fig. \ref{fig:infections_japan_1dim} shows the number of daily new infections and the total number of them versus time in Japan in comparison with the one-dimensional theoretical curves of Eqs.(\ref{eq:avrami_sol_j_1dim}) and (\ref{eq:avrami_sol_d_1dim}). The 95\%CI was calculated as same as that in Fig. \ref{fig:ln1div1_x_t2_japan} and shown by the broken lines along the theoretical curves. The parameters of the theoretical curves were adjusted to achieve a close fit between the equation and the $D(t) - t$ curve by the least-square scheme using the slopes shown in Fig. \ref{fig:ln1div1_x_t2_japan}. The $J(t) - t$ curve has larger relative 95\%CI range, e.g. $509 \pm84 (17\%)$ at the maximum $J(t)$, than the $D(t) - t$ curve, e.g. $5657 \pm92 (1.6\%)$ at the maximum $J(t)$ simply because J(t) is the number of daily announce of the PCR-tested-positive persons and $D(t)$ is the sum of them. The obtained values of the parameter were $D_s = 15150$ [person], $0.3780 - 0.0047 \leq S_c G N_0 \leq 0.3780 + 0.0042$ [1/day], $\nu = 0.0090$ [1/day], where $D(t)$ was the value obtained by subtracting the sample value on March 26, 2020 as the baseline value and $\nu$ was assumed to be constant.

\begin{figure} [t]
  \centering
  \includegraphics[width=8.0cm]{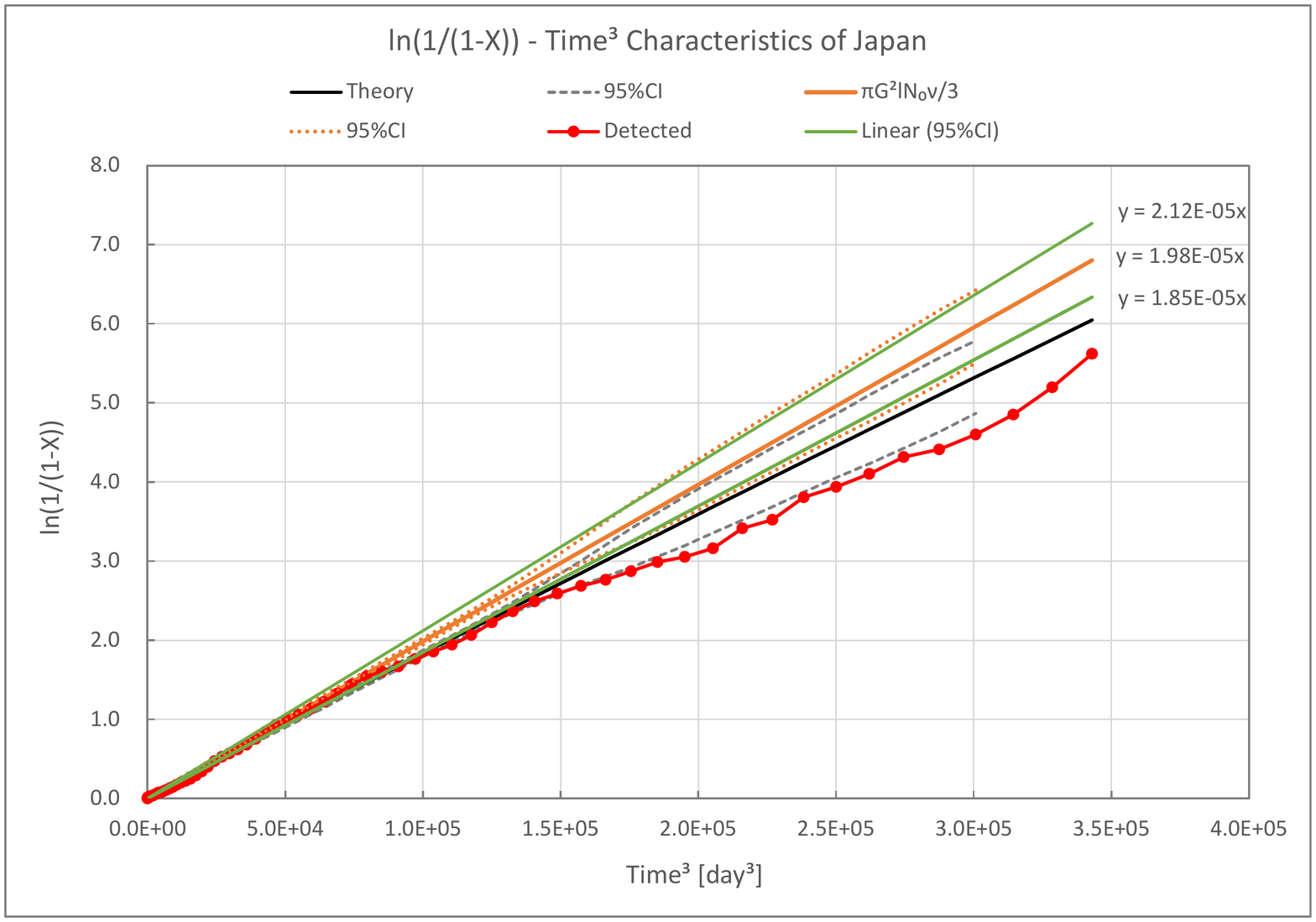}
  \caption{$\mathrm{ln}(1/(1-X)) – Time^3$ characteristics of Japan. The red marker represents the detected infections, the black line the theoretical curve (2-dim.), the gray broken lines the 95\%CI, the orange line the theoretical line (2-dim.) for the small $\nu t$ region, the orange dotted lines the 95\%CI, and the green lines linear approximations of the 95\%CI curves.}
  \label{fig:ln1div1_x_t3_japan}
\end{figure}

Using the same data, and assuming the two-dimensional growth, Fig. \ref{fig:ln1div1_x_t3_japan} shows the $\mathrm{ln}(1/(1-X)) – Time^3$ characteristics of Japan similar to Fig. \ref{fig:ln1div1_x_t2_japan}. The 95\%CI was calculated as same as that in Fig. \ref{fig:ln1div1_x_t2_japan}. The  slope of the orange line was estimated to be $1.98 \times 10^{-5} - 1.33 \times 10^{-6} \leq slope \leq 1.98 \times 10^{-5} +1.37 \times 10^{-6}$ as shown in the figure, which gives the value of $\pi G^2 l_c N_0 \nu /3$ in Eq.(\ref{ln1div1_x_sol_s_nt_2dim}).

\begin{figure} [t]
  \centering
  \includegraphics[width=8.0cm]{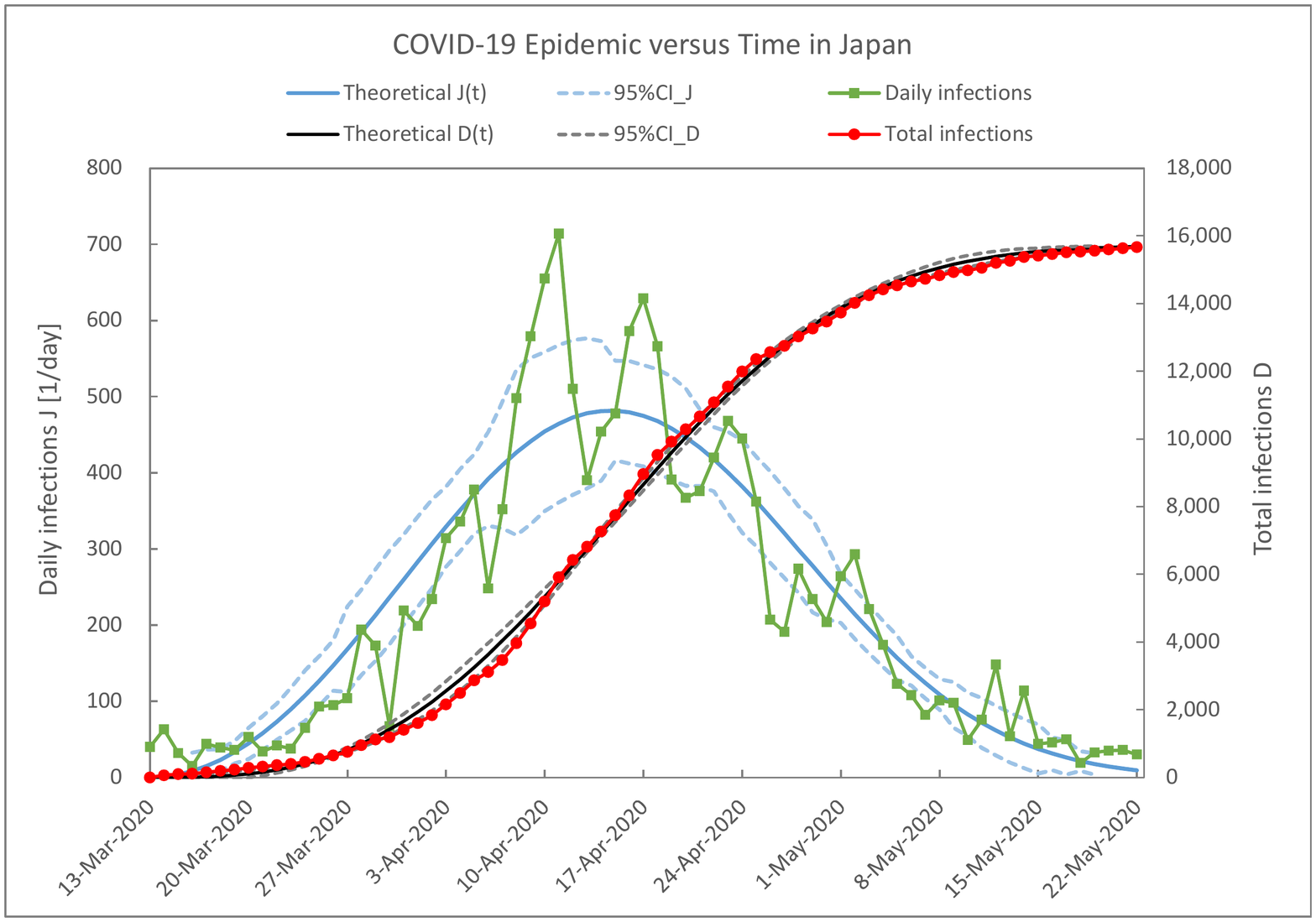}
  \caption{COVID-19 epidemic versus time in Japan. The red marker represents the detected total infections, the black line the theoretical curve (2-dim.) of $D(t)$, the gray broken lines the 95\%CI, the green marker the detected daily infections, the blue curve the theoretical curve (2-dim.) of $J(t)$, and the blue broken lines the 95\%CI curves.}
  \label{fig:Patients_Japan_2dim}
\end{figure}

Fig. \ref{fig:Patients_Japan_2dim} shows the number of daily new infections and the total number of them versus time in Japan in comparison with the two-dimensional theoretical curves of Eqs.(\ref{eq:avrami_sol_j_2dim}) and (\ref{eq:avrami_sol_d_2dim}). The 95\%CI was calculated as same as that in Fig. \ref{fig:ln1div1_x_t2_japan}. The parameters of the theoretical curves were obtained similarly to the one-dimension case, $D_s = 15720$ [person], $0.0170 - 0.0011 \leq 2 \pi G^2 l_c N_0 \leq 0.0170 + 0.0012 \mathrm{ [1/day^{2}]}, \nu = 0.0070$ [1/day], where $D(t)$ was the value obtained by subtracting the sample value on March 13, 2020 as the baseline value and $\nu$ was assumed to be constant. 

The one-dimensional model was superior to the two-dimensional model because the standard deviations of $\mathrm{ln}(1/(1-X))$ data were 0.043 and 0.252, and those of $D(t)$ were 135 and 217, respectively.

\subsection{COVID-19 data of Tokyo}

\begin{figure} [p]
  \centering
  \includegraphics[width=8.0cm]{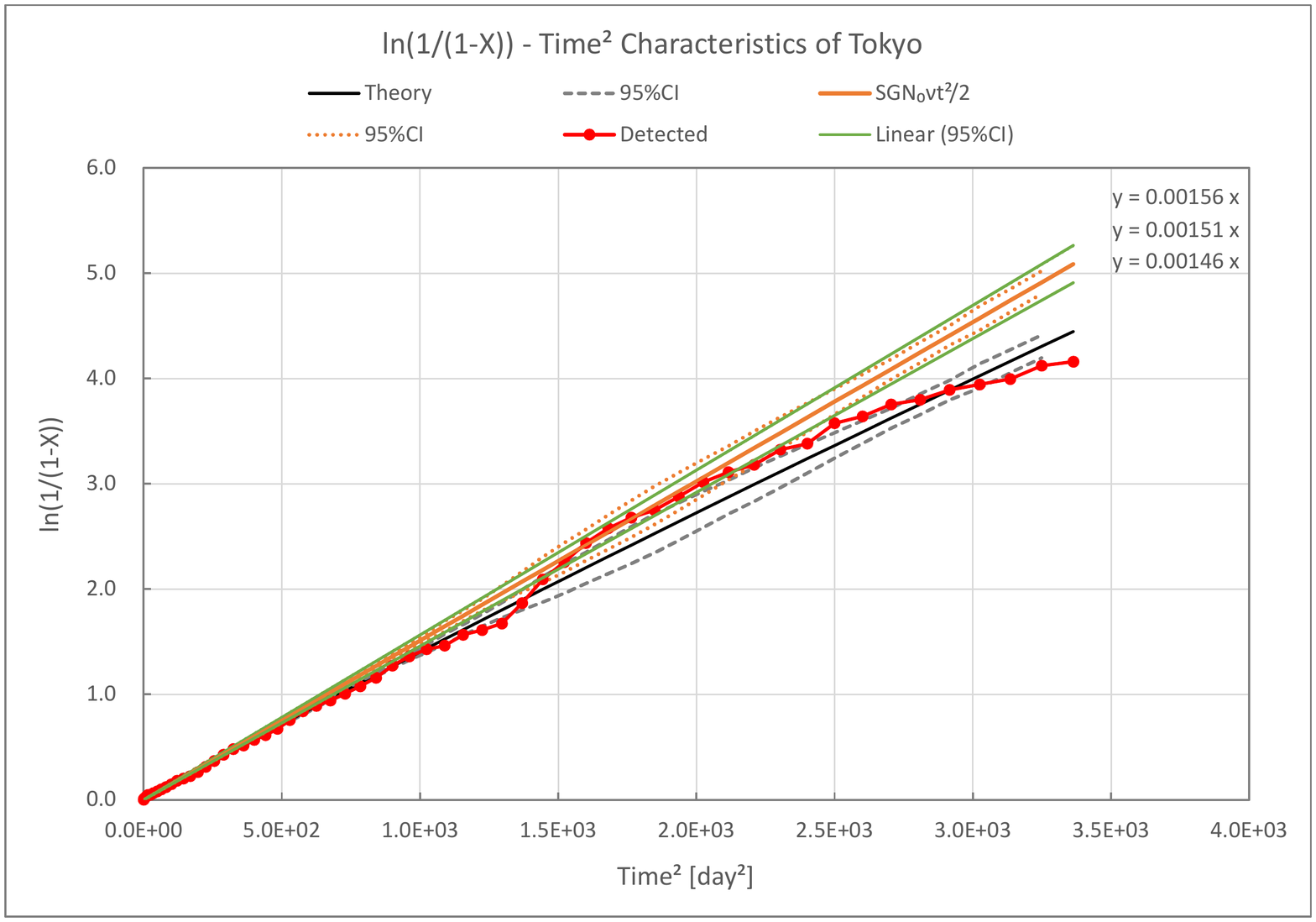}
  \caption{$\mathrm{ln}(1/(1-X)) – Time^2$ characteristics of Tokyo. The red marker represents the detected infections, the black line the theoretical curve (1-dim.), the gray broken lines the 95\%CI, the orange line the theoretical line (1-dim.) for the small $\nu t$ region, the orange dotted lines the 95\%CI, and the green lines linear approximations of the 95\%CI curves.}
  \label{fig:ln1div1_x_t2_tokyo}
\end{figure}

The COVID-19 data of Tokyo is provided in some web sites. Using the data\cite{infections_tokyo}, and assuming the one-dimensional growth, Fig. \ref{fig:ln1div1_x_t2_tokyo} shows the $\mathrm{ln}(1/(1-X)) – Time^2$ characteristics of Tokyo similar to Fig. \ref{fig:ln1div1_x_t2_japan}. The slope of the orange line was estimated to be $0.00151 \pm 0.00005$ as shown in the figure.

\begin{figure} [p]
  \centering
  \includegraphics[width=8.0cm]{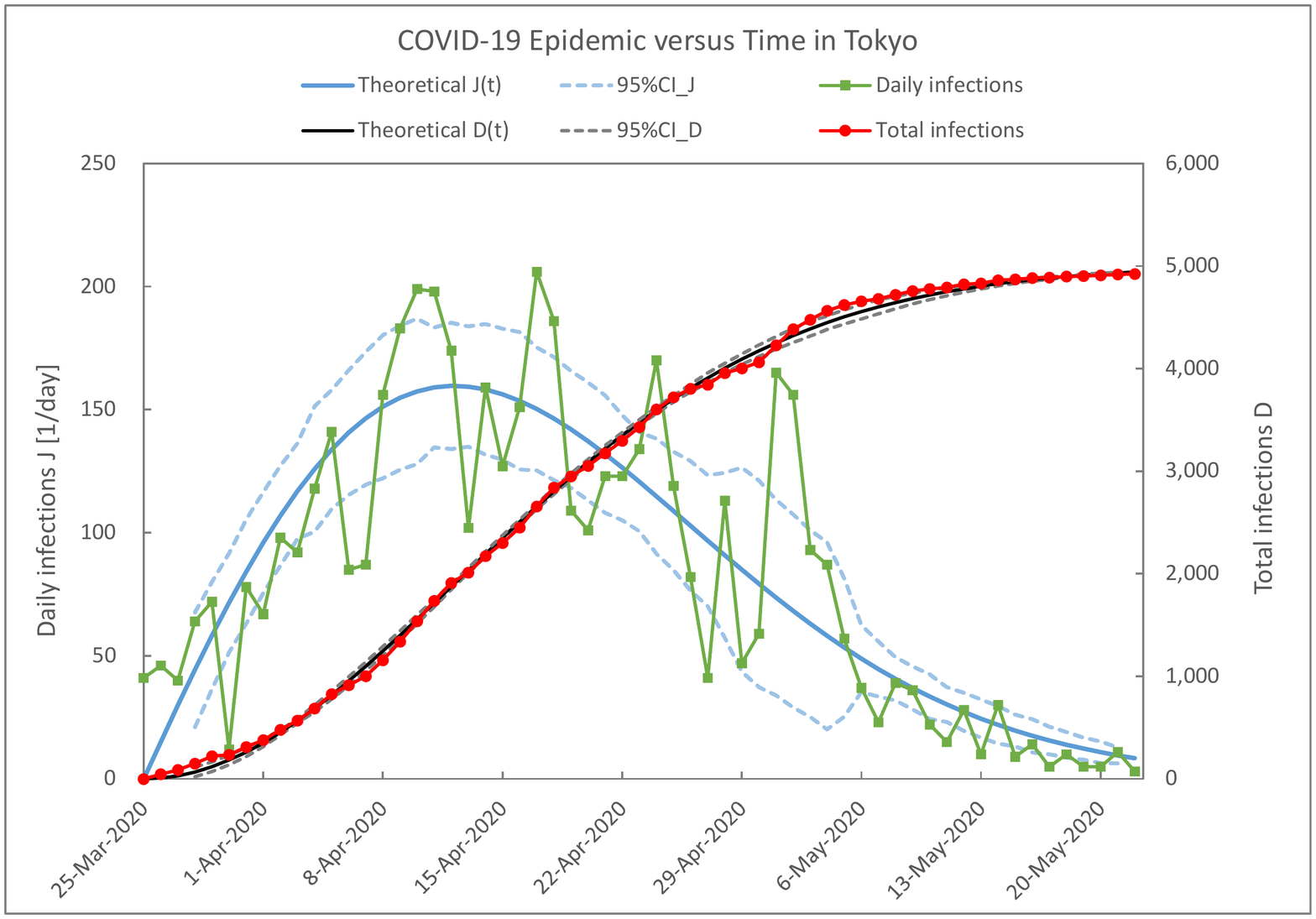}
  \caption{COVID-19 epidemic versus time in Tokyo. The red marker represents the detected total infections, the black line the theoretical curve (1-dim.) of $D(t)$, the gray broken lines the 95\%CI, the green marker the detected daily infections, the blue curve the theoretical curve (1-dim.) of $J(t)$, and the blue broken lines the 95\%CI curves.}
  \label{fig:infections_tokyo_1dim}
\end{figure}

Fig. \ref{fig:infections_tokyo_1dim} shows the number of daily new infections and the total number of them versus time in Tokyo in comparison with the one-dimensional theoretical curves of Eqs.(\ref{eq:avrami_sol_j_1dim}) and (\ref{eq:avrami_sol_d_1dim}). The parameters of the theoretical curves were obtained as mentioned above, $D_s = 5000$ [person], $0.4200 - 0.0144 \leq S_c G N_0 \leq 0.4200 + 0.0133$ [1/day], $\nu = 0.0072$ [1/day], where $D(t)$ was the value obtained by subtracting the sample value on March 25, 2020 as the baseline value and $\nu$ was assumed to be constant.

\begin{figure} [p]
  \centering
  \includegraphics[width=8.0cm]{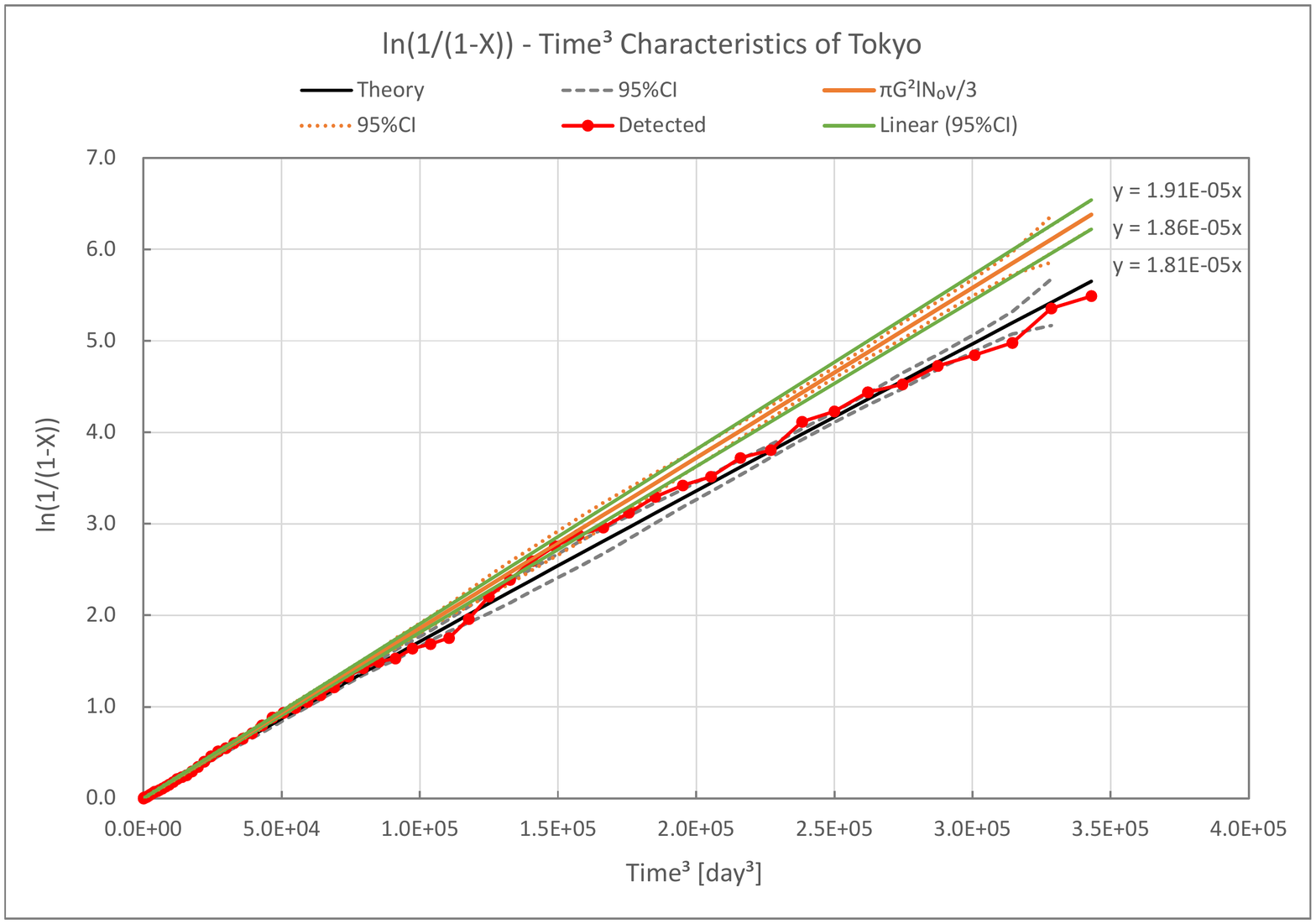}
  \caption{$\mathrm{ln}(1/(1-X)) – Time^3$ characteristics of Tokyo. The red marker represents the detected infections, the black line the theoretical curve (2-dim.), the gray broken lines the 95\%CI, the orange line the theoretical line (2-dim.) for the small $\nu t$ region, the orange dotted lines the 95\%CI, and the green lines linear approximations of the 95\%CI curves.}
  \label{fig:ln1div1_x_t3_tokyo}
\end{figure}

Using the same data, and assuming the two-dimensional growth, Fig. \ref{fig:ln1div1_x_t3_tokyo} shows the $\mathrm{ln}(1/(1-X)) – Time^3$ characteristics of Tokyo similar to Fig. \ref{fig:ln1div1_x_t3_japan}. The slope of the orange line was estimated to be $1.86 \times 10^{-5} \pm 5.0 \times 10^{-7}$ as shown in the figure.

\begin{figure} [p]
  \centering
  \includegraphics[width=8.0cm]{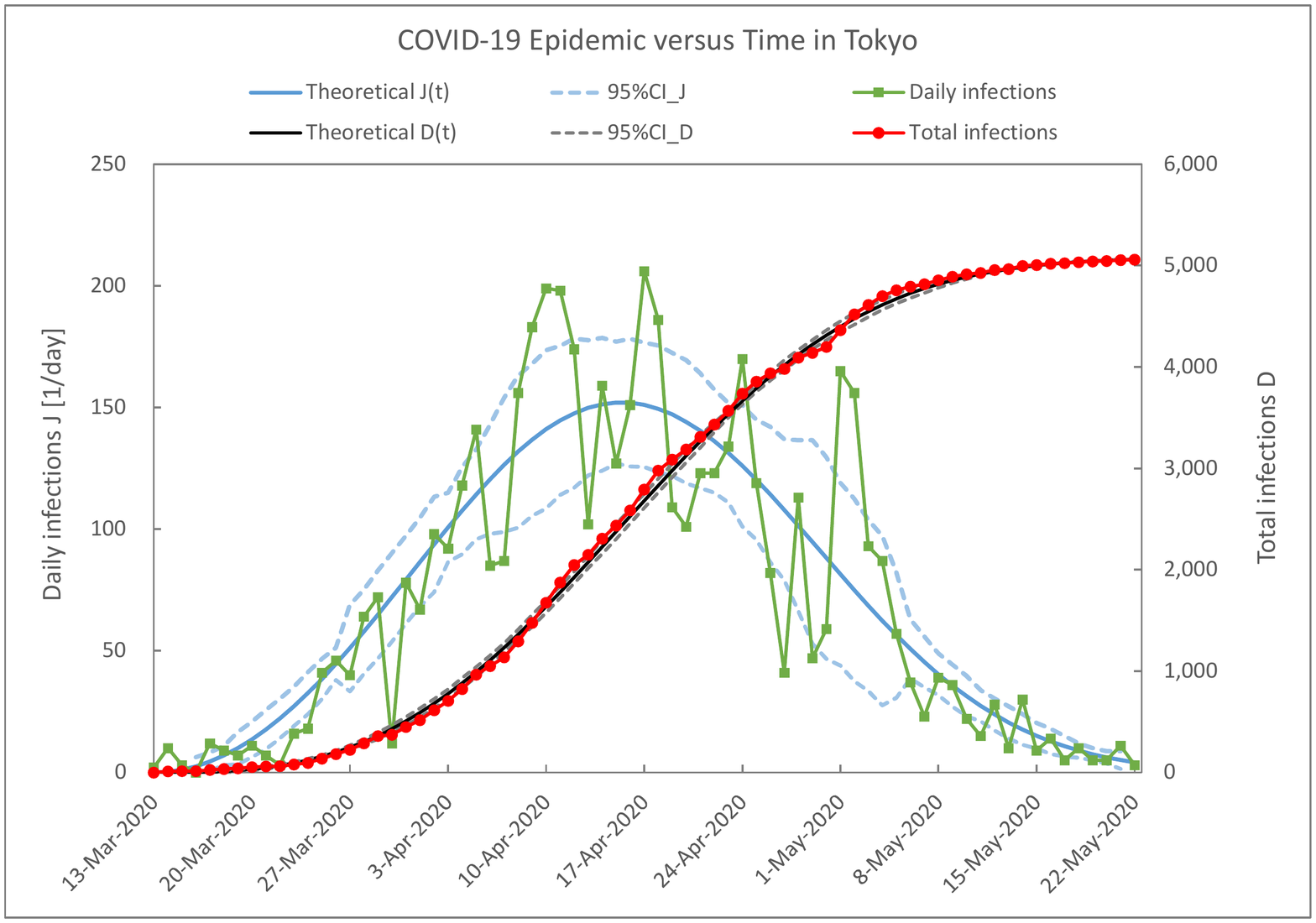}
  \caption{COVID-19 epidemic versus time in Tokyo. The red marker represents the detected total infections, the black line the theoretical curve (2-dim.) of $D(t)$, the gray broken lines the 95\%CI, the green marker the detected daily infections, the blue curve the theoretical curve (2-dim.) of $J(t)$, and the blue broken lines the 95\%CI curves.}
  \label{fig:infections_tokyo_2dim}
\end{figure}

Fig. \ref{fig:infections_tokyo_2dim} shows the number of daily new infections and the total number of them versus time in Tokyo in comparison with the two-dimensional theoretical curves of Eqs.(\ref{eq:avrami_sol_j_2dim}) and (\ref{eq:avrami_sol_d_2dim}). The parameters of the theoretical curves were obtained as mentioned above, $D_s = 5080$ [person], $0.0155 - 0.0004 \leq 2 \pi G^{2} l_c N_0 \leq 0.0155 + 0.0004 \mathrm{[1/day^{2}]}$, $\nu = 0.0072$ [1/day], where $D(t)$ was the value obtained by subtracting the sample value on March 13, 2020 as the baseline value and $\nu$ was assumed to be constant. 

The one-dimensional model was a little inferior to the two-dimensional one because the standard deviations of $\mathrm{ln}(1/(1-X)$ were 0.119 and 0.081, respectively, although those of $D(t)$ were 55 and 55, respectively.

\section{Discussion}

In order for one new phase to form in the basic phase, a new phase must be born and grow. In the case of COVID-19, the nucleus is the first person who causes successive infection. The location of nucleation changes randomly as the infected person moves. If the nucleus is born at each place, it spreads in one or two dimensions. This process fits the random nucleation and growth model. The characteristics have been determined by three parameters: $D_s$, domain growth rate $S_c G N_0$ or $2 \pi G^2 l_c N_0$, and $\nu$. 

In Japan, there were approximately 16,000 presumed susceptible between March 13 and May 24, 2020. This is only about 0.13\% of the total Japanese population of 12,616,000. It corresponds to the limited domain growth area rather than the large number of immunes because  the antibody positive rate was announced on June 16, 2020 to be 0.10\% in Tokyo and 0.17\% in Osaka\cite{antibody_test} or 0.43\%\cite{antibody_test_softbank}.

The value of the one-dimensional growth parameter $S_c G N_0$ which is proportional to the growth speed $G$, was a small constant of 0.378 [1/day] in Japan and 0.420 [1/day] in Tokyo. The inverse of these values, 2.7 and 2.4 days for Japan and Tokyo, respectively, may be proportional to their serial intervals. The value of $\nu$ was a small constant of 0.0090 [1/day] in Japan and 0.0072 [1/day] in Tokyo. When the half-life $t_{1/2}$ of the remained nuclei is calculated similar to the radioisotope, for example, $t_{1/2} = \ln 2/0.0090 \simeq 77$ days, which is considered to relate to a period of the epidemic. 

In Japan, the contact chance with infected person and the range of random movement of the infected are estimated to be small since February 2020. Many Japanese began to pay attention to COVID-19 in February 2020. Experts have taken measures against epidemic clusters since February 2020. It is considered that such behavior became a factor that restricted the nucleation and growth. There are views that there may be other immunological factors, which are expected to be verified scientifically.

\section{Conclusion}
An attempt was made to analyze the transition of the number of newly infected persons with COVID-19 in Japan from March 13 to May 24, 2020 by the dynamic phase transition theory. As a result, the epidemic was in good agreement with the random nucleation followed by the one- or two- dimensional linear growth basic model. The epidemic in Japan in that period was simply rate-limited by three parameters that could be regarded as constant, the initial susceptible, domain growth rate, and nucleation decay constant. This model provides a good predictor of epidemic because it consists of one equation and the initial $\mathrm{ln}(1/(1-X))$ specific plot is linear.

\end{document}